\documentclass{artikel3}
\usepackage[T1]{fontenc}
\usepackage[portuges,english]{babel}
\usepackage{amssymb,latexsym,amsmath,color}
\usepackage[colorlinks,linkcolor=blue,urlcolor=blue,citecolor=black,
plainpages=false,pdfpagelabels,breaklinks]{hyperref}
\usepackage{palatino}
\usepackage{mathpazo}

\title{Contradiction, Quantum Mechanics, and the Square of Opposition}
\author{{Jonas R. Becker Arenhart}\thanks{Department of Philosophy, Federal University of Santa Catarina, Florianópolis, SC 88040-900, Brazil (jonas.becker2@gmail.com).} \and {D\'ecio Krause}\thanks{Department of Philosophy, Federal University of Santa Catarina, Florianópolis, SC 88040-900, Brazil.}}
\begin{document}
\maketitle

\newcommand{\ita}{\textit}

\begin{abstract}
We discuss the idea that superpositions in quantum mechanics may involve contradictions or contradictory properties. A state of superposition such as the one comprised in the famous Schrödinger's cat, for instance, is sometimes said to attribute contradictory properties to the cat: being dead and alive at the same time. If that were the case, we would be facing a revolution in logic and science, since we would have one of our greatest scientific achievements showing that real contradictions exist. We analyze that claim by employing the traditional square of opposition. We suggest that it is difficult to make sense of the idea of contradiction in the case of quantum superpositions. From a metaphysical point of view the suggestion also faces obstacles, and we present some of them.
\medskip

\end{abstract}

\section{Introduction}

The peculiar way superposition figures in quantum mechanics is sometimes said to be ``the'' greatest feature of the theory. It brings such a level of novelty and conceptual embarrassment that a proper understanding of this phenomenon is still lacking, despite the efforts by many distinct interpretations advanced over the years. Superposition is involved in most of the typical quantum enigmas such as Schrödinger's cat and the associated measurement problem; it is also central to the Einstein-Bohr debate and to the problem of identical particles. Furthermore, superposition is responsible for part of the great empirical applications and technological innovations based on the theory.

In a recent paper, da Costa and de Ronde \cite{cos13} suggested that superpositions may be better understood if a paraconsistent approach could be adopted: their claim is that assuming that a superposition involves contradictions may be fruitful for our understanding of superpositions and of quantum mechanics in general. Now, while the idea that superpositions involve contradictions is a rather new one, this is not the first time in the history of quantum theory that a contradiction is said be found in the core of the theory. Recall the early attempts at understanding the atom with Bohr's model, employing both classical mechanics and quantum principles to describe the atom. Bohr's model of the atom is generally claimed to be inconsistent. Furthermore, the wave-particle duality and Bohr's Complementarity Principle are sometimes said to involve contradictions, or at least to generate paradoxical situations. In a broadest context, contradictions are not rare in science; they are said to appear in the early formulations of the infinitesimal calculus, in Frege's \ita{Grundgezetse} and in Cantor's naive set theory, among other contexts. So, the suggestion that superpositions may involve contradictions is no singular situation in the history of science.\footnote{See da Costa and French \cite{cos03} and Vickers \cite{vic13} for further discussions on inconsistency in science.}

Roughly put, the idea that superposition is inconsistent is very simple: a state $\Psi$ in a superposition $\Psi = \alpha\phi_1 + \beta\phi_2$ (with $\alpha^2 + \beta^2 = 1$) corresponds to a state simultaneously attributing contradictory properties to the system represented by this state. Obviously, this idea may be understood as endorsing the system with actual contradictory properties or, alternatively, as attributing it only a contradictory property in a ``possible realm''. However the ontological status of contradictory properties is interpreted, consider in particular a spin-$\frac{1}{2}$ system in the sate $|\uparrow_z\rangle$. It is generally accepted without controversy that this system has spin up in the $z$ direction. But consider another direction, like $x$. In order to determine the probabilities of obtaining the spin in $x$ as up or down, we have to convert the state of the system from $|\uparrow_z\rangle$ to a superposition of $|\uparrow_x\rangle$ and $|\downarrow_x\rangle$. Now, it is the system in such a superposition that is said to have \ita{contradictory properties}, because the projectors $|\downarrow_x\rangle\langle\downarrow_x|$ and $|\uparrow_x\rangle\langle\uparrow_x|$, representing the respective properties of having spin down and spin up in the $x$ direction, represent in fact \ita{contradictory properties}. So, quantum mechanics would be committed with contradictions somehow, and a paraconsistent logic would be the most natural candidate for helping our understanding of such situations.

Here we shall examine precisely the suggestion that superpositions involve contradictions. In particular, in section \ref{cont} we shall suggest --- keeping with the case of the $\frac{1}{2}$-spin system in the state $|\uparrow_z\rangle$ and further examples --- that properties represented by projectors like $|\downarrow_x\rangle\langle\downarrow_x|$ and $|\uparrow_x\rangle\langle\uparrow_x|$ are perhaps not best understood as contradictory, so that if our suggestion is sound, superpositions of the corresponding eigenstates would not really involve contradictions. In section \ref{square} we apply the square of opposition to this case and advance the thesis that such states may represent rather \ita{contrary} propositions. We suggest that it is possible to provide for an interpretation in order to extend plausibly the square to the hexagon of opposition (for further information on the hexagon of opposition, see \cite{bez12}). In section \ref{met} we shall investigate the suggestion at a metaphysical level. Our suggestion is that even though a metaphysics of contradictory properties is not objectionable in principle, some difficulties appear in connection with quantum mechanics. We conclude with some remarks concerning the relation of logic and ontology in quantum mechanics.

\section{Contradictions in quantum mechanics}\label{cont}

Now, let us take a closer look at the thesis that superpositions may involve contradictions. As originally proposed by da Costa and de Ronde \cite{cos13}, the idea involves two separated assumptions: first of all, for a $\frac{1}{2}$-spin system in the state $|\uparrow_z\rangle$, properties represented by projectors $|\downarrow_x\rangle\langle\downarrow_x|$ and $|\uparrow_x\rangle\langle\uparrow_x|$ are contradictory. That is, properties ``to have spin up in the $x$ direction'' and ``to have spin down in the $x$ direction'' would be contradictory  with one another. Secondly, the state of superposition of the two states $|\uparrow_x\rangle$ and $|\downarrow_x\rangle$ would be a contradiction, it attributes both (contradictory) properties to the system.

Obviously, the description of the previous paragraph takes into account only a particular system in a particular state, but the idea is that the thesis according to which there may be contradictions in quantum mechanics can be generalized for any superposition of states $s_1$ and $s_2$ that are classically incompatible (da Costa and de Ronde also call them ``classically inconsistent'' --- we shall investigate the meaning of this statement later), so that a superposition of them provides for contradictory attribution of properties. In particular, it seems that the famous double-slit experiment also involves contradictions: the state $\Psi = \psi_1 + \psi_2$, with $\psi_1$ meaning that the particle went through the first slit, and $\psi_2$ meaning that it went through the second slit should also be seen as attributing it contradictory properties. Schrödinger's cat, in a state $\Phi = C_d + C_a$, where $C_d$ means that the cat is dead and $C_a$ means that it is alive is also supposed to be contradictory. We shall consider these cases in the course of our exposition.

Our main interest in this section concerns the sense of ``contradiction'' that is at stake here (we shall not provide an exhaustive analysis of this concept). We shall suggest that even though it is tempting to see contradictions in superpositions, the idea may be resisted, so that perhaps this is not a genuine case of inconsistency in science.

As traditionally understood, a contradiction is a conjunction of a pair of propositions $A$ and $B$ such that one is the negation of the other, which  are called \ita{contradictories}. Obviously, to state a contradiction one must employ a language with negation, and natural language and many other artificial languages in logic and mathematics satisfy this requirement (this is a syntactical requirement). A second important feature of a contradiction is encapsulated by the square of opposition (see \cite{bez12}), and concern the semantics of a contradiction: in supposing classical logic, when $A$ and $B$ are contradictory, one of them is true and the other false, we cannot have both true or both false (this is the semantic requirement). Let us encapsulate those requirements on a language $\pounds$ as follows:
\begin{enumerate}
\item The language $\pounds$ must have a negation sign;
\item Contradictory statements of $\pounds$ must always have opposite truth values.
\end{enumerate}

Given those features of a contradiction, are we able to investigate whether a superposition like $\alpha|\uparrow_x\rangle + \beta|\downarrow_x\rangle$ represents a contradiction in the traditional sense. In order for that to happen, at least the following two conditions must be assured, for our previous requirements for a contradiction to obtain:
\begin{enumerate}
\item $|\uparrow_x\rangle$ and $|\downarrow_x\rangle$ must both obtain and also assumed to be the negation of one another.
\item The statements corresponding to properties represented by $|\downarrow_x\rangle\langle\downarrow_x|$ and $|\uparrow_x\rangle\langle\uparrow_x|$ must have opposite truth values.
\end{enumerate}

Consider now the first case. To begin with, by accepting for the sake of argument that the states $|\uparrow_x\rangle$ and $|\downarrow_x\rangle$ could somehow be taken as one being the negation of the other (so that the syntactical requirement could be satisfied), we should also accept that the plus sign $+$ in Hilbert spaces (where we express  a superposition as a sum of vectors) represents some form of conjunction, of ``obtaining together''. In the case of the double slit, we would have that the system $\Psi = \psi_1 + \psi_2$ would be describing a situation in which a particle goes by both slits. For the Schrödinger cat, the state $\Phi = C_d + C_a$ describes a situation in which the cat is both dead and alive. Now, the problem is that hardly anyone would allow that the sum could be read as conjunction, since their meanings are very different. If the distinction is not immediate, one can check this from the very fact that they are different operations, there are some properties distinguishing them: conjunction is idempotent, while sum of vectors in a Hilbert space is not. So, there is a first difficulty in passing from superposition to contradiction, as usually understood, for a vector sum does not immediately amounts to simultaneous reality (whatever this means).

Now, besides that point concerning the differences between vectors sums and conjunctions, let us examine the idea that $|\uparrow_x\rangle$ and $|\downarrow_x\rangle$ are states to which contradictory properties are attributed. In order for that to happen, we would have to allow that some feature of Hilbert space representation, in this case, orthogonalization,  represents negation, and second, that their semantic features fit the role of contradictory statements as mentioned above. For ease of reading, let us call $u_x$ the proposition that the system has spin up in the $x$ direction; $d_x$ is the proposition that the system has spin down in the $x$ direction. Then, according to the suggestion above, $u_x$ and $d_x$ are contradictory. Let us investigate this proposal.

If it is indeed true that $u_x$ and $d_x$ are contradictory, then both cannot be true and both cannot be false at once. Let us see whether this is the case in quantum mechanics. To begin with, let us state a minimal condition for property attribution, one which will also make room for modal interpretations and the paraconsistent one we are examining here, and which shall guide part of our investigations (for this minimal condition, see also Muller and Saunders \cite[p.513]{mul08}):

\medskip
\begin{description}
\item[The minimal property attribution condition:] If a system is in an eigenstate of an operator with eigenvalue \textbf{v}, then the system has the qualitative property corresponding to such value of the observable.
\end{description}

\medskip
Notice that this is weaker than the usual eigenvalue-eigenvector link, stating only a conditional. The interpretation we are examining here proposes to complement the minimal condition by describing how a system bears properties when not in an eigenstate:

\medskip
\begin{description}
\item[Paraconsistent property attribution:] When in a superposition, the system does have the properties related to the vectors forming the superposition, and they are contradictory.
\end{description}

\medskip
Let us check how those two principles attribute properties to quantum systems and whether there are reasons to suppose that $u_x$ and $d_x$ are contradictory (recall the semantic condition that must be fulfilled). First of all, when the system has spin up in the $x$ direction, \ita{i.e.}, when the statement $u_x$ is true, then by the above minimal condition, the probability of having spin up is $1$ and the probability of having spin down in the $x$ direction is $0$, so that it is false that the system has spin down in the $x$ direction, \ita{i.e}, $d_x$ is false. The same holds when $d_x$ is true: we have that $u_x$ is false. So, when the conditions for application of the minimal principle are met, both states have opposite truth values. But the job is still not done: we must still grant that one of those propositions must always be the case (being so that the other one will be false), as the semantic requirement for a contradiction seems to demand. In the case of spin for $\frac{1}{2}$-systems, this corresponds to the requirement that for any given spatial direction, one (and only one) of the two options ``up'' or ``down'' will always have to be the case. However, as we shall suggest, there are difficulties in satisfying this requirement along accepting the paraconsistent property attribution.

First of all, it seems that the semantic requirement that one of the two terms in a superposition must always be the case (so that we can have a contradiction) is in fact in conflict with the paraconsistent property attribution principle. For the latter principle to apply, in the case of a superposition, both ``up'' and ``down'' would \ita{have to be the case simultaneously}. Recall what happens in the case of the two slit or Schrödinger's cat: according to this proposal, the particle must go by both slits, the cat must be dead and alive. So, there cannot be alternate truth values in this case, for both must be simply true. So, there is a conflict of the paraconsistent property attribution principle with the very requirement that the vectors in a superposition stand for contradictory properties, at least according to the usual semantic requirements related to contradictions, as it appears in the traditional analysis of this concept.\footnote{We do not think that traditional definitions are untouchable and sacred, but if a proposal is supposed to violate traditional standards, then it is those traditional standards that must be taken for granted and shown to be violated.} It seems that one cannot have both the claim that $u_x$ and $d_x$ are contradictory and the claim that a superposition involves contradictions, as supplied by the paraconsistent property attribution principle. As it stands, it seems, these demands are incompatible.

In the second place, there seem to be problems with both (incompatible) conditions even when taken on their own. To begin with, it is not always plausible to suppose that for any property, a quantum system must be so that it has the qualitative property corresponding to one of the associated eigenvalues, while the other ones do not obtain. That is, the semantic requirement for a contradiction, when applied to statements such as $d_x$ and $u_x$ supposed contradictories, seems to be in conflict with most interpretations of quantum mechanics and even with some of its limitative theorems. As is well-known, by results such as the Kochen-Specker theorem, one cannot attribute truth values `truth' and `false' to every proposition corresponding to property attribution to a system, so that it is not always the case that for any property whatever, a system either has it or does not have it. Given that quantum mechanics is a probabilistic theory and that one cannot attribute only $0$s and $1$s as probability for the possible outcomes of measurements, some properties are not allowed to have such a comfortable feature of being instantiated or not instantiated. Furthermore, this requirement seems to lead to problems in the results in the case of some experiments. For instance, consider the two slit experiment. If it is indeed the case that the statements ``the particle went through slit 1'' and ``the particle went through slit 2'' are contradictory and have always opposite truth values, then the expected result of a two slit experiment would never be the interference pattern typical of the experiment, but rather the pattern obtained when the slits are opened one at a time alternatively.

Furthermore, the paraconsistent property attribution principle also seems to generate problems of its own (we shall discuss this point again in section \ref{met}). It is not clear that, as in the case of a two slit experiment, the attribution of both properties to the system will generate the interference patterns. Rather, it seems more plausible to suppose that superposition is a phenomenon not reducible to property attribution, even if it is a paraconsistent one. Furthermore, the paradox of the cat that is dead or alive does not get closer to being understood when we suppose that it has both properties. So, once again, attributing both properties to the system does not seem to generate a better understanding of the phenomenon. Finally, this kind of move makes it even more difficult to understand how a typical measurement of a system in superposition yields always determinate results, but not contradictory results: one must be able to explain how a property possessed by the system disappears, while the other one remains.

So, as we shall suggest in the next section, states in superposition do not seem to involve contradictions. Perhaps the fact that there is not a straightforward way to read a contradiction from a superposition can be also seen from the formal analysis employed in da Costa and de Ronde \cite[p.855]{cos13}, where a paraconsistent set theory $ZF_1$ is employed. Considering a system S which is in a superposition of states $s_1$ and $s_2$, the authors introduce a predicate symbol $K(S, s_1)$ to represent the predicate that ``S has the superposition predicate associated with $s_1$''. The same reading holds with obvious adaptation for $K(S, s_2)$ and for $\neg K(S, s_1)$ and $\neg K(S, s_2)$. To account for a contradiction, a \ita{Postulate of Contradiction} is introduced: when S is in a superposition of $s_1$ and $s_2$, we have $$K(S, s_1) \wedge \neg K(S, s_1) \wedge K(S, s_2) \wedge \neg K(S, s_2).$$ ``This means that superposition implies contradiction'' \cite[p.855]{cos13}.

Notice that in this postulate the contradiction does not come from the fact that two states $s_1$ and $s_2$ form the superposition or are related somehow, but rather from the assumption that in a superposition the conjunction $K(S, s_1) \wedge \neg K(S, s_1)$ is present (as well as $K(S, s_2) \wedge \neg K(S, s_2)).$ That is, the contradiction comes not from a relation between $s_1$ and $s_2$, but from a relation of $s_1$ with its negation (which obviously is a contradiction in the language; the same holds for the case of $s_2$). Now, that seems to change the locus of the contradiction; it is placed no longer in the statement that properties represented by projectors like $|\downarrow_x\rangle\langle\downarrow_x|$ and $|\uparrow_x\rangle\langle\uparrow_x|$ are contradictory and it seems hard to say that under this particular analysis $s_1$ and $s_2$ (or, for our previous example, $u_x$ and $d_x$) are contradictory. The introduction of a negation sign to allow for a contradiction is a symptom of that change of focus. It seems now that in a superposition $s_1$ is contradictory with its own negation, and the same holds for $s_2$.

However, that move seems to be unmotivated. Everything that a superposition puts in a system, in terms of the vectors that are part of the superposition, this reading also takes off from the superposition, so that in the end we have no information at all. In the case of the two slit experiment, the system $\Psi = \psi_1 + \psi_2$ is indeed stating, according to this reading, that the system went through the first slit, it did not went through the first slit, it went through the second slit, and went not through the second slit. It seems that the contradiction cancels the informational charge of a superposition.

Also, concerning the role of the negation in the above formal analysis, some issues arise. It seems that there are two options concerning the role of the negation sign here: either $\neg K(S, s_1)$ is equivalent to $K(S, s_2)$ or it is not. In the second case, then, as we mentioned previously, the contradiction comes from postulation of both $K(S, s_1)$ and $\neg K(S, s_1)$, not from any peculiar relation between $s_1$ and $s_2$ in superposition. In this case, the claim is at odds with the suggestion that the vectors forming a superposition are contradictory among themselves. On the other hand, if there is an equivalence, then it should be a theorem or another axiom of the paraconsistent theory $ZF_1$ that $K(S, s_1) \leftrightarrow \neg K(S, s_2)$ and also $K(S, s_2) \leftrightarrow \neg K(S, s_1)$. Our suggestion is that such equivalence would become problematic for systems whose basis can be composed by more than two vectors. Consider now a simple system in superposition of three orthogonal states $a$, $b$ and $c$, and for simplicity, let $p_a$, $p_b$ and $p_c$ represent the propositions ``the system is in $a$'', ``the system is in $b$'' and ``the system is in $c$'', respectively. Now, since each proposition is equivalent to the negation of the others, we have i) $p_a \leftrightarrow \neg p_b$, ii) $p_c \leftrightarrow \neg p_a$ and iii) $p_c \leftrightarrow \neg p_b$. By a chain of equivalences, from iii) and ii) we have $p_c \leftrightarrow \neg p_a$, and from this result and i) we have $p_a \leftrightarrow \neg p_a$. Of course, that may be fine for when we use a paraconsistent logic, but seems to be the a strange analysis when a system is indeed in the state $p_a$: that would imply that it is also not in such a state. So, once again, this does not seem to be the correct analysis of the contradiction, and the contradiction does not result from a relation from the states in superposition, but rather is introduced by the Postulate of Contradiction.

\section{The square of opposition}\label{square}

Given that ``contradiction'' is not really mandatory on us in cases of superpositions, are there any alternatives for us to understand logically the relation of states in superposition? Even more generally, how are we to understand the relation between quantum observables such as $|\downarrow_x\rangle\langle\downarrow_x|$ and $|\uparrow_x\rangle\langle\uparrow_x|$? We suggest that the square of opposition may be used to throw some light on this issue (see also \cite{bez03}, \cite{bez12}). Let us present the details of our analysis.

Recall that the square was introduced in medieval times to represent the logical relationships between the four categorical propositions in Aristotelian logic and to capture some imediate inferences holding among them. The four propositions are $A$: ``All $S$ are $P$'', $E$: ``No $S$ is $P$'', $I$: ``Some $S$ are $P$'', and $O$: ``Some $S$ are not $P$. Their relations of opposition are summarized as follows:
\begin{enumerate}
\item $A$ and $E$ are contrary.
\item $A$ and $O$ as well as $E$ and $I$ are contradictory.
\item $I$ and $O$ are subcontrary.
\item $I$ is subaltern of $A$, and $O$ is subaltern of $E$.
\end{enumerate}

Having said that, we shall not use categorical propositions, but rather the propositions $u_x$ and $d_x$ presented before as our case study, and enlarge the idea of the square to cope with these ``propositions'' as well. That is, we shall investigate their logical relations according to the frame of the square of opposition. To begin with, we suggest that $u_x$ and $d_x$ are better thought as \ita{contraries} rather than as contradictories: contraries are propositions that cannot both be true, but that can both be false. Indeed, if it is true that $d_x$, then obviously it is false that $u_x$, and if it is true that $u_x$, then it is false that $d_x$. So, in our opinion quantum mechanics is not contradictory after all when a superposition is written. Then, at the place of the square of opposition where one usually finds propositions $A$ and $E$ we have $u_x$ and $d_x$, bearing the same logical relation.

An immediate problem appears now: given that $u_x$ and $d_x$ are contraries and not contradictories, what could be the contradictory of $u_x$? And what would be the contradictory of $d_x$? For that question to be answered, obviously, we would have to introduce a negation sign. For that purpose, we introduce the negation of a quantum proposition, allowing that the contradictory of $u_x$ would be $\neg u_x$, the proposition expressing that ``the system does not have spin up in the $x$ direction''. Then, it is clear that besides $u_x$ and $d_x$ being contrary, $u_x$ and $\neg u_x$ are contradictory, as well as $d_x$ and $\neg d_x$. This gives us what traditionally is pictured as the relation that $A$ bears with $O$ and that $E$ bears with $I$, respectively.

Notice that from a quantum mechanical point of view the negation sign introduced here presents no problem. It is true in the theory that a system either has a given property (have spin up in the $x$ direction, for instance) or does not have it. The real novelty comes from the distinct ways a system may \ita{fail} to have a property: by being in a superposition, by not being in the required eigenstate because its state is an eigenstate of an observable that does not commute with the observable being measured, and perhaps even more.

The definition above of contradictory quantum sentences also gives us immediately that $\neg d_x$ is subaltern to $u_x$, and $\neg u_x$ is subaltern to $d_x$. Obviously, if the system has spin up in $x$, then it does not have spin down in $x$, and correspondingly, if the system has spin down in $x$, then, it is not up in $x$. So subalterns are also granted by our schema. Following the traditional notation, this accounts for the relation between $A$ and $I$ on the one hand, and between $E$ and $O$ on the other.

The only thing left to complete the square is a proper relation of subcontrariety. According to our definition, $\neg u_x$ and $\neg d_x$ would be the candidates to be subcontraries of one another. Do they satisfy the requirement, namely, can they be both true, but not both false? First, they can both be true: a system may neither have spin up nor down in a given direction, being it enough that it is in a superposition of up and down in that direction. Can both be false? Not really, because that would give us that the system is up and down in the same direction, which is impossible. This accounts for the traditional relation between $I$ and $O$.

To accomplish an even better relation with the situation in quantum mechanics, however, we still can complete the square to obtain an hexagon. As usually understood (again, see \cite{bez12}), the hexagon complements the the traditional square by adjoining a top element, understood as totality, the disjunction from $A$ and $E$, and a bottom element, the conjunction of $I$ and $O$. Let us see that they make complete sense in our schema for the quantum case.

For the case of the top, we must have one of both $d_x$ and $u_x$, making their disjunction. This is nice to represent a situation where there is only epistemic ignorance as to the state of the system. That is, we don't know whether the system is up or down, but it is surely in one of those states, and a measurement of the respective observable allow us to determine which is the case. One of them will be the outcome of a possible measurement. Don't confuse it with superposition. For the bottom case, when we have both $\neg d_x$ and $\neg u_x$, it can be used to represent a typical situation of a system in a state in a superposition: before the measurement, it is neither up nor down. This is precisely the opposite of what is proposed by the paraconsistent property attribution principle, but it seems to respect the intuition that quantum superpositions are phenomena not reducible to classical phenomena, in particular, not to classical predication. We shall present more difficulties for the paraconsistent approach in the next section, and we hope they also motivate our reading of superposition.

\section{Contradictory metaphysics and quantum mechanics}\label{met}

Now, we change focus and concentrate on the idea that a superposition may be accepted as incorporating contradictory properties just like systems in an eigenstate do have properties. Even if the candidate property represented by the superposition is assumed to exist only at a possible or potential level, we suggest that some difficulties appear.

The very idea that some mathematical piece employed to develop an empirical theory may furnish us information about unobservable reality requires some care and philosophical reflection. The greatest difficulty for the scientifically minded metaphysician consists in furnishing the means for a ``reading off'' of ontology from science (see Arenhart and Krause \cite{ar14} for a related discussion). What can come in, and what can be left out? Different strategies may provide for different results, and, as we know, science does not wear its metaphysics on its sleeves. The first worry may be making the metaphysical piece compatible with the evidence furnished by the theory.

The strategy adopted by da Costa and de Ronde \cite{cos13} may be called top-down: we investigate higher science and, by judging from the features of the objects described by the theory, we look for the appropriate logic to endow it with just those features. In this case (quantum mechanics), there is the theory, apparently attributing contradictory properties to entities, so that a logic that does cope with such feature of objects is called forth.\footnote{A similar approach is used in the case of the problem of identity and individuality of particles in quantum mechanics: according to some views, identity and individuality are lost for quantum particles. Non-reflexive logics are logics developed to reflect that fact (see French and Krause \cite{fre06}), and are strong enough to accommodate the development of a non-reflexive version of quantum mechanics, that is, the theory is built again with the respective metaphysical principles already embedded in the underlying logic (see \cite{dom08}).} Now, even though we believe that this is the in great measure the right methodology to pursue metaphysics within scientific theories, there are some further methodological principles that also play an important role in these kind of investigation, principles that seem to lessen the preferability of the paraconsistent approach over alternatives.

To begin with, let us focus on the paraconsistent property attribution principle. According to that principle, recall, the properties corresponding to the vectors in a superposition are all attributable to the system, they are all real. The first problem with this rendering of properties (whether they are taken to be actual or just potential) is that such a superabundance of properties may not be justified: as Maudlin \cite{mau13} has recalled (although in a different context), not every bit of a mathematical formulation of a theory needs to be reified. Some of the parts of the theory are just that: mathematics required to make things work, others may correspond to genuine features of reality. The greatest difficulty is to distinguish them, but we should not assume that every bit of it corresponds to an entity in reality. So, on the absence of any justified reason to assume superpositions as a further entity on the realms of properties for quantum systems, we may keep them as not representing actual properties (even if merely possible or potential ones).

That is, when one takes into account other virtues of a metaphysical theory, such as economy and simplicity, the paraconsistent approach seems to inflate too much the population of our world. In the presence of more economical candidates doing the same job and absence of other grounds on which to choose the competing proposals, the more economical approaches take advantage. Furthermore, considering economy and the existence of theories not postulating contradictions in quantum mechanics, it seems reasonable to employ Priest's razor --- the principle according to which one should not assume contradictions beyond necessity (see Priest \cite{pri87}) --- and stick with the consistent approaches. Once again, a useful methodological principle seems to deem the interpretation of superposition as contradiction as unnecessary.

The paraconsistent approach could take advantage over its competitors, even in the face of its disadvantage in order to accommodate such theoretical virtues, if it could endow quantum mechanics with a better understanding of quantum phenomena, or even if it could add some explanatory power to the theory. In the face of some such kind of gain, we could allow for some ontological extravagances: in most cases explanatory power rules over matters of economy. However, it does not seem that the approach is indeed going to achieve some such result. As we mentioned before, attributing contradictory properties to a quantum system seems to simply eliminate the superposition as a source of information. Nothing is said about the system, because everything said is also denied.

Besides that lack of additional explanatory power or enlightenment on the theory, there are some additional difficulties here. There is a complete lack of symmetry with the standard case of property attribution in quantum mechanics. As it is usually understood, by adopting the minimal property attribution principle, it is not contentious that when a system is in one eigenstate of an observable, then we may reasonably infer that the system has the property represented by the associated observable, so that the probability of obtaining the eigenvalue associated is $1$. In the case of superpositions, if they represented properties of their own, there is a complete disanalogy with that situation: probabilities play a different role, a system has a contradictory property attributed by a superposition irrespective of probability attribution and the role of probabilities in determining measurement outcomes. In a superposition, according to the proposal we are analyzing, probabilities play no role, the system simply has a given contradictory property by the simple fact of being in a (certain) superposition.

For another disanalogy with the usual case, one does not expect to observe a system in such a contradictory state: every measurement gives us a system in particular state, never in a superposition. If that is a property in the same foot as any other, why can't we measure it? Obviously, this does not mean that we put measurement as a sign of the real, but when doubt strikes, it may be a good advice not to assume too much on the unobservable side. As we have observed before, a new problem is created by this interpretation, because besides explaining what is it that makes a measurement give a specific result when the system measured is in a superposition (a problem usually addressed by the collapse postulate, which seems to be out of fashion now), one must also explain why and how the contradictory properties that do not get actualized vanish. That is, besides explaining how one particular property gets actual, one must explain how the properties possed by the system that did not get actual vanish.

Furthermore, even if states like $\frac{1}{\sqrt{2}}(|\uparrow_x\rangle + |\downarrow_x\rangle)$ may provide for an example of a candidate of a contradictory property, because the system seems to have both spin up and down in a given direction, there are some doubts when the distribution of probabilities is different, in cases such as $\frac{2}{\sqrt{7}}|\uparrow_x\rangle + \sqrt{\frac{3}{7}}|\downarrow_x\rangle$. What are we to think about that? Perhaps there is still a contradiction, but it is a little more inclined to $|\downarrow_x\rangle$ than to $|\uparrow_x\rangle$? That is, it is difficult to see how a contradiction arises in such cases. Or should we just ignore the probabilities and take the states composing the superposition as somehow opposed to form a contradiction anyway? That would put metaphysics way too much ahead of science, by leaving the role of probabilities unexplained in quantum mechanics in order to allow a metaphysical view of properties in.

\section{Final remarks}

To finish these remarks on the paraconsistent approach to superposition, we would like to emphasize that this core quantum concept (superposition, and in particular, entanglement), still seems to lack an adequate logical foundation. As Paul Dirac has said (\cite[p.12]{dir11}), it cannot be understood by means of classical concepts, and we guess that we need even another kind of logic to deal with it (we have proposed a particular kind of modal logic for that; see Krause, Arenhart and Merlussi \cite{kra14}), but it seems to us that this logic is not a paraconsistent one. The idea that superposition may be understood in terms of contradiction and paraconsistent logic still requires some elaboration in order to overcomes some of the doubts the proposal naturally raises.

That is, even though we are sympathetic to the use of alternative logics to deal with ontological problems in science and with the foundations of science itself, it seems that a paraconsistent approach to quantum mechanics is still facing many difficulties; more articulation is still required in order for the approach to work. In the face of the compatibility of distinct metaphysical packages with quantum mechanics, some of them more economic than the paraconsistent one, one should develop and further clarify the paraconsistent view in order to be able to clearly present its advantages over the alternatives. As usual, attributions of contradiction to some physical theory does face the problem of making the contradiction explicit, and we hope that our analysis with the square of oppositions has made it clear that in some cases quantum propositions are indeed in some kind of opposition, but the opposition is contrariety, not contradiction proper.

Also, we do not believe that the approach to quantum propositions by the square as suggested above is easily generalizable to quantum propositions of any sort. In particular, for observables with continuous spectra it seems that it is not possible to deal with quantum propositions taking just two at each time and studying their relations as to what kind of opposition obtains. If this is really the case, then, obviously, it is still in complete agreement with our claim throughout the paper that superposition is a completely \ita{sui generis} phenomenon.

\end{document}